# Vulnerability Management for an Enterprise Resource Planning System

Shivani Goel
Assistant Professor
Computer Science and
Engineering Department
Thapar University Patiala
Punjab India-147001

Ravi Kiran
Professor
School of Behavioral Sciences
and Business Studies
Thapar University Patiala
Punjab India-147001

Deepak Garg
Associate Professor Computer
Science and Engineering
Department
Thapar University Patiala
Punjab India-147001

## ABSTRACT

Enterprise resource planning (ERP) systems are commonly used in technical educational institutions(TEIs). ERP systems should continue providing services to its users irrespective of the level of failure. There could be many types of failures in the ERP systems. There are different types of measures or characteristics that can be defined for ERP systems to handle the levels of failure. Here in this paper, various types of failure levels are identified along with various characteristics which are concerned with those failures. The relation between all these is summarized. The disruptions causing vulnerabilities in TEIs are identified .A vulnerability management cycle has been suggested along with many commercial and open source vulnerability management tools. The paper also highlights the importance of resiliency in ERP systems in TEIs.

## 1. INTRODUCTION

ERP systems are used for providing an approach that integrates the functioning of various departments in any technical educational institution (TEI), to conduct the operations from a central database with accuracy and convenience. Any ERP system involves an information system which has two parts: hardware and software. Hardware part can be called as infrastructure system which includes network, databases and computer peripherals including servers. The software part is the data and information which flows through hardware. There can be failures in both parts. The hardware and software resources might fail and information may get corrupted and it can lead to ERP failure. Various types of failure conditions are handled differently in systems ensuring security. These systems can have different characteristics defined for handling failure and hence security. These are discussed in next section.

## 2. FAILURE CONDITIONS

During the life cycle of a system, there occur certain situations which cause many threats to system security. The system is required to have certain security attributes in that situation. This relationship between conditions and attributes is depicted in figure 1.

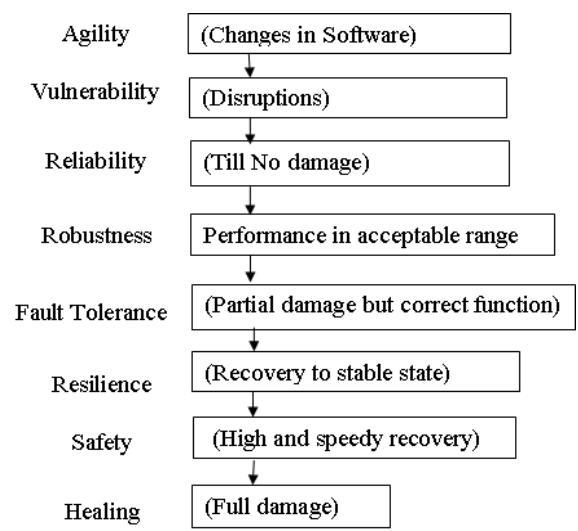

**Figure 1 : Relation between failure conditions and attributes**

With no disruptions, the system works accurately. Whenever there is a need for change, disruptions can occur in a system which can cause harm to system operations. The probability of disruptions is judged by vulnerability of the system. The system should be least vulnerable in order to predict as many disruptions as possible. A reliable system is supposed to maintain correct functioning of the system till there is no damage. A system should be highly reliable so that it can perform without damage for a longer duration. In case there is no damage, but certain noise, the robust system will be able to perform within an acceptable range of errors. Thus, a system needs to be highly robust so that it can perform near correct functions even in the presence of noise. When there is partial damage to the system, but the system is still performing the correct function, the system is said to be fault tolerant. But only fault tolerance will not makeup for the partial damage to the system. So the system needs to recover back to stable state while performing correct function. Thus a system should be resilient so that it can recover as early as possible to a stable state after a partial damage. In other words, resilience engineering goes beyond reliability engineering and robust engineering [1]. For this, the system needs to have flexibility so that it can recover early. The capacity to predict the disruptions should be as high as possible so that it can take measures to avoid or recover fast. Recovery may need resources, so sustainability is required which can be provided at a high level with greater amount of redundancy. Sustainability is defined as the system's property that measures the balanced generation and consumption of the system resource and supply resources when needed. If a system is able to speedy recovery, it is said to be safe . In case a system gets full damage, i.e. resilience fails, the system needs to be healed. Healing is defined as Healing refers to recovery of a lost function of the system (after damage) by means of the external resource instead of a system's own resource. Thus, the main source of system failure is vulnerabilities. The earliest the system can identify the vulnerabilities, better will be the security of the system.

## 3. DISRUPTIONS FOR ERP IN TEIs

The main reasons for failures are disruptions caused in a system. In order to assess the security of an ERP system, the common disruptions leading to vulnerabilities are needed to be identified. A total of 25 technical users from TEIs were consulted using unstructured interviews as an information gathering tool for identifying the major disruptions in ERP in TEIs. The major technical disruptions identified in ERP system in a TEI are network failure, failure of surveillance system and interface issues. These can cause vulnerabilities like attack on the web based interface for getting the database configuration management, attack on network devices, wireless access points etc. Poor security system, inadequate training and poor error tracking system have been identified as organizational factors. Due to gap in security of the network, attackers can get access to network's critical assets. Poor error attacking system can cause serious attacks unnoticed which can consume system resources. The error by the person operating the ERP system can also cause the system to fail. The wrong intensions of the users in ERP may cause malicious attacks in the system like man in the middle attack, phishing and spoofing attacks etc . Various sources of disruptions which may cause vulnerabilities in ERP in TEIs are shown in figure 2 below:

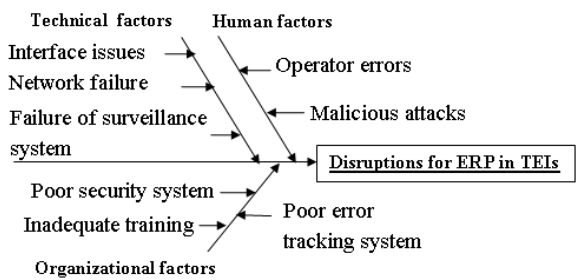

**Figure 2: Disruptions for ERP in TEIs**

ERP can be best suited for handling vulnerabilities caused by disruptions when it supports resilience. This is because a resilient system is capable of supporting recovery mechanisms along with correct functioning when it finds vulnerabilities. Resilience of a system is measured by the level of its vulnerability to a specific risk [2, **4 to** 3]. Reducing vulnerability has a positive impact on the resilience of a system [**13 to** 4]. Resilience depends upon the anticipating unexpected disruptive events and designing solutions to eliminate errors as early as possible [**9 to** 5]. The attributes of a resilient system are highlighted in next section.

## 4. ATTRIBUTES OF A RESILIENT SYSTEM

There are many attributes which a resilient system should have. These are identified as adaptive capacity, agility, fault tolerance , flexibility and redundancy [6]. Each of these is defined as follows:

a) **Adaptive capacity** Adaptability demonstrates the ability to adapt to changing environments while delivering the intended functionality under changing operating conditions.

b) **Agility** Agility has been used in conjunction with flexibility as a defining attribute of resilience [7]. According to Christopher and Peck [3], resilience involves agility and it helps a system to rapidly re-organise itself. Morello [8], on the other hand, suggests that agility may introduce new risks and vulnerabilities which result in lower resilience.

c) **Fault Tolerance** Fault tolerance is defined as the ability to deliver service in the presence of faults. The ability to function after software component damage under attack is a kind of the fault tolerance in software and is the resilience of a software system [9].

d) **Flexibility** Fiksel [10] defines flexibility as a major system characteristic that contributes to resilience. Helaakoski [11] relates flexibility to agility and adaptability, which indicates that flexibility is a system's ability to rapidly adapt to its changing environment.

e) **Redundancy** Redundancy is defined as keeping extra capacity or resources kept in reserve to be used in case of a disruption [4]. According to Haimes et al. [12], redundancy is the ability of certain components of a system to support the functions of failed components without any considerable effect on the performance of the system itself.

From the analysis of all these definitions, it is found that an ERP system should support flexibility and redundancy to be a resilient system. Since vulnerability detection is the key to success for a resilient system, it is important to handle vulnerabilities in the system. A vulnerability management cycle is proposed in next section.

## 5. VULNERABILITY MANAGEMENT

Vulnerability management is a pro-active approach for managing security of a network because it involves continuous monitoring of system conditions for identifying vulnerabilities. Regular assessment of vulnerability is important for checking the security system of any system. Various users have studied various methods for ensuring security in ERP systems. A multi layer tree model for enterprise vulnerability management has been proposed by Wu and Wang [13] which helps in qualifying the overall score of the company. A vulnerability management cycle has been proposed which consists of following activities:

### 5.1.1 Planning for known vulnerability

Before starting with any management, a planning is required. There are a large number of resources and processes which can be attacked by vulnerabilities. When any part is affected by the attack, then that needs to be cured. But for a pro-active approach, it is impossible to monitor all of these. In order to identify vulnerabilities, the resources or processes which are important and are susceptible to vulnerabilities must be first identified. The type of potential threats to each identified resource must be listed. This will lead to a plan for monitoring for the threats becoming a reality. The known types of vulnerabilities can be listed with their symptoms while for the unknown types some methods can be learned only from organizational data and experience. Examples identified for ERP in TEIs are man in the middle attack, denial of service attack,

### 5.1.2 Monitoring for vulnerability

Regular monitoring of the system activities identified in first step need to be done continuously. This can be done by regular network scanning, firewall logging, penetration testing . There is also a tool called vulnerability scanner which can be used for the purpose.

### 5.1.3 Analyzing to identify vulnerabilities

This involves analyzing the results indicated during the monitoring of the vulnerability. If there are symptoms for some threat becoming a reality, then mitigation for that vulnerability should be used to minimize the consequences if an attack occurs.

### 5.1.4 Mitigating the vulnerabilities

The process of finding how to prevent the vulnerabilities. Patches can be applied in the affected area. The vendors of the affected software or hardware can provide the patch as early as possible to minimize the ill effects of the vulnerability attack.

### 5.1.5 Updating the list of known vulnerabilities

In the case of new vulnerabilities found, the information should be updated so that the planning can include for newly identified vulnerabilities for monitoring in future.

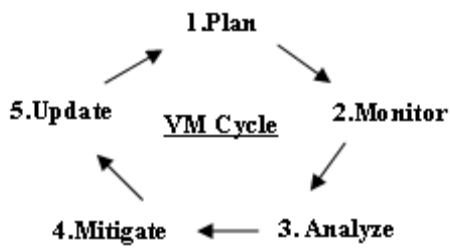

**Figure 3: Vulnerability management cycle**

The above cycle can be used in any system. In order to implement it, the vulnerabilities need to be identified first so that planning for those can be done. In this paper, the vulnerabilities are identified in an ERP system in TEIs. In any ERP system, there is a requirement for security at many levels since there are many areas where vulnerabilities can occur. The most common is in authentication and authorization procedures. This is because due to centralized systems, a number of users need to be assigned different roles and responsibilities. But in case of lack of proper security, there is a threat of wrong manipulation of information. The man in the middle attack, SQL injections, data theft, least privileges violations are common. This can be monitored by network scanner tools. The errors caused due to operator errors also cause disruptions in the system. Proper training should be provided to reduce operator errors. Another mitigating action is providing the Autocorrect facilities to handle the errors due to wrong entry by operators. In order to avoid these, it is advisable to introduce some security solutions at the nearest points of vulnerability attacks. This will also ease the vulnerability analysis for identifying the attacks as early as possible. There are many tools available in the market for various processes in vulnerability management. These are either available as open source and are free in the market or these are commercial. Some of these are shown in figure 4:

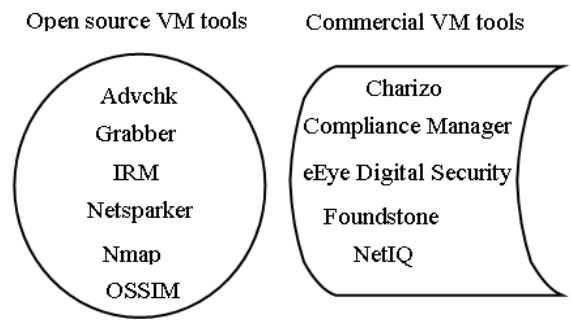

**Figure 4: Tools for Vulnerability Management**

1. Advchk (Advisory Check), gathers security advisories automatically and compares them to a list of known services, and gives an alert if you are vulnerable.

2. Grabber, A scanner which analyses vulnerabilities in web applications.

3. Information Resource Manager (IRM) a powerful Web-based asset tracking .

4. Mavutina Netsparker, aweb application security scanner.

5. Nmap is a free, open source utility for network exploration or security auditing.

6. OSSIM (Open Source Security Information Management) which is used to provide a network/security administrator a detailed view of the network and devices.

Examples of commercial vulnerability management tools are:

1. Attachmate's NetIQ Vulnerability Manager enables users to define and maintain configuration policy templates, vulnerability bulletins, and automated checks via AutoSync technology.

2. eEye Digital Security: It provides a suite consists of the Retina Network Security Scanner which is a vulnerability assessment tool, Blink Professional which is a host-based security technology, and the REM Security Management Console.

3. Mayflower Chorizo Intranet edition , a scanner for intranet web applications

4. McAfee's Foundstone Enterprise is also solution that offers asset discovery, inventory, and vulnerability prioritization with threat intelligence, correlation, remediation tracking, and reporting and it also does not use agents.

5. Symantec BindView's Compliance Manager is a software-based solution which allows organizations to evaluate their assets against corporate standards or industry best practices, without the need for agents.

Along with these tools, security metrics such as vulnerability counts, intrusion attempts, unauthorized access attempts can also be used for monitoring vulnerabilities in ERP systems for TEIs[14].

## 6. CONCLUSION

For any ERP system, agility is required to be an essential feature to be successful. Allowing changes at any time increases the probability of disruptions and hence failures. So if there is flexibility in a system to allow changes, the system should be flexible enough to incorporate recovery mechanisms also [3,4,15]. The crucial condition is when there is partial damage in the system and speedy recovery is required so that system is safe. This indicates the importance of resilience. In order to enhance resilience, adaptive capacity and hence flexibility should be increased even after a disruption. The redundancy within the system can help increase sustainability of a system and hence can pave the way to faster recovery before full damage[10]. Resilience is a system's ability to bounce back from disruptions and disasters by building in redundancy and flexibility[16]. Thus, some form of redundancy is found to be important for ERP system in TEIs. In any ERP system for TEIs, failure or disruptions like network failure, operator error and malicious attacks on data are inevitable. The ability to predict these disruptions can be one step in taking preventive measures for handling these before these become a reality. This enhances the security of the system. Vulnerability management cycle can be used for monitoring and predicting the vulnerabilities in the system. Various tools can be used for handling vulnerabilities.